\documentclass[10pt,english,prd,aps,showpacs,epsf,floats]{revtex4}
\usepackage[T1]{fontenc}
\usepackage[latin1]{inputenc}
\usepackage{amsmath}
\usepackage{amssymb}
\usepackage{amsfonts}
\usepackage{babel}

\makeatletter
\makeatletter
\input{tcilatex}
\makeatother
\makeatother

\begin{document}

\title{\textbf{Local softening of information geometric indicators of chaos
in statistical modeling in the presence of quantum-like considerations}}
\author{Adom Giffin$^{1}$, Sean A. Ali$^{2,3}$, Carlo Cafaro$^{4,5}$}
\affiliation{$^{1}$Department of Mathematics and Computer Science, Clarkson University,
13699 Potsdam, New York, USA }
\affiliation{$^{2}$International Institute for Theoretical Physics and Mathematics
Einstein-Galilei, 59100 Prato, ITALY}
\affiliation{$^{3}$Department of Arts and Sciences, Albany College of Pharmacy and Health
Sciences, 12208 Albany, New York, USA}
\affiliation{$^{4}$Max-Planck Institute for the Science of Light, 91058 Erlangen, GERMANY }
\affiliation{$^{5}$Institute of Physics, Johannes Gutenberg University Mainz, 55128
Mainz, GERMANY}

\begin{abstract}
In a previous paper (C. Cafaro \textit{et al}., 2012), we compared an
uncorrelated $3D$ Gaussian statistical model to an uncorrelated $2D$
Gaussian statistical model obtained from the former model by introducing a
constraint that resembles the quantum mechanical canonical minimum
uncertainty relation. Analysis was completed by way of the information
geometry and the entropic dynamics of each system. This analysis revealed
that the chaoticity of the $2D$ Gaussian statistical model, quantified by
means of the Information Geometric Entropy (IGE), is \textit{softened} or
weakened with respect to the chaoticity of the $3D$ Gaussian statistical
model, due to the accessibility of more information. In this companion work,
we further constrain the system in the context of a correlation constraint
among the system's micro-variables and show that the chaoticity is further
weakened, but only \textit{locally}. Finally, the physicality of the
constraints is briefly discussed, particularly in the context of quantum
entanglement.
\end{abstract}

\maketitle

\section{Introduction}

In the paper by L. A. Caron \textit{et al}. \cite{caron1}, classical chaos
is compared with quantum chaos, and the authors discuss why the former is%
\textit{\ weaker}\ than the latter. It was suggested that the weakness of
quantum chaos may arise from quantum fluctuations that give rise to
Heisenberg's uncertainty relation. It is also known that a quantum
description of chaos is qualitatively different from a classical description
and that the latter cannot simply be considered an approximation of the
former. In fact, the only aspect of quantum theory that may be retained by a
corresponding classical description is the canonical Heisenberg's
uncertainty relation, specifically, a minimum spread of order $\hbar ^{n}$
in the $2n$-dimensional phase space \cite{peres} (where $\hbar \overset{%
\text{def}}{=}\frac{h}{2\pi}$ and $h$ is Planck's constant).

In a previous paper \cite{cafarosoft}, we studied the information geometry
and the information-constrained dynamics of a $3D$ uncorrelated Gaussian
statistical model and compared it with that of a $2D$ uncorrelated Gaussian
statistical model, which was obtained from the higher-dimensional model via
introduction of an additional information constraint that resembled the
Heisenberg uncertainty relation. We showed that the chaoticity (temporal
complexity) of the $2D$ uncorrelated Gaussian statistical model
(quantum-like model), quantified by means of the Information Geometric
Entropy (IGE) \cite{cafaroPD} and the Jacobi vector field intensity, was 
\textit{softened} relative to the chaoticity of the $3D$ uncorrelated
Gaussian statistical model (classical-like model). By \textit{softened}, we
mean any attenuation in the asymptotic temporal\ growth of the indicators of
chaoticity. It is worth noting that the statistical models in question were
limited to the extent that we assumed that the correlation between the
micro-variables of the system was unknown.

In this paper, we will again discuss the manner in which the degree of
complexity changes for a statistical model (the probabilistic description of
a physical system) in the presence of incomplete knowledge when the
information-constrained dynamics, the so-called entropic dynamics \cite%
{catichaED}, on the underlying curved statistical manifolds becomes even
more constrained. Furthermore, we will reduce the probabilistic description
of the dynamical systems in the presence of partial knowledge to information
geometry (Riemannian geometry applied to probability theory, see \cite{amari}%
) and inductive inference \cite{AThesis, c2-1,c2-2, c4}. We employ the same
theoretical framework developed for this, termed the \emph{Information
Geometric Approach to Chaos }(IGAC) \cite{cafarothesis, cafaroCSF}, where
information geometric techniques are combined with maximum relative entropy
methods \cite{AThesis, c2-1,c2-2, c4} to study the complexity of
informational geodesic flows on curved statistical manifolds (statistical
models) underlying the probabilistic description of physical systems in the
presence of incomplete information. We expand our previous findings by
further constraining the quantum-like $2D$ uncorrelated model, herein
denoted as $2Du$, with knowledge of the correlation between the microscopic
degrees of freedom of the system by way of a covariance term, $\sigma _{xy}$%
. Our analysis not only provides evidence that the degree of chaoticity of
statistical models is related \linebreak to the existence of uncertainty
relation-like information constraints, as was seen before,\ it also
demonstrates that the chaoticity is also dependent upon the covariance term
parameterized in terms of a correlation coefficient.

This constraint, specifically the correlation coefficient, may well have a
physical interpretation. It is known, for example, that a realistic approach
to generate entangled quantum systems is via dynamical interaction, of which
local scattering events (collisions) are a natural, ubiquitous type \cite%
{Bubhardt}. Indeed, we have shown in a recent work \cite{KACM} how the IGAC
can be used to examine the quantum entanglement of two spinless,
structureless, non-relativistic particles, where the entanglement is
produced by two Gaussian wave-packets interacting via a scattering process.
In that work, it was shown how the correlation coefficient can be related to
two-particle squeezing\ parameters\ \cite{Serafini} for the case of
continuous variable quantum systems with Gaussian continuous degrees of
freedom \cite{peter}.

The layout of this article is as follows. In Section 2, we present the basic
differential geometric properties of the quantum-like two-dimensional
uncorrelated statistical model, $2Du$, and the further constrained version,
namely, the two-dimensional correlated statistical model, herein denoted as $%
2Dc$. In\ Section 3, we describe the geodesic paths on the curved
statistical manifolds underlying the entropic dynamics of the two
statistical models. In Sections 4 and 5, we study the chaotic properties of
the information-constrained dynamics on the underlying curved statistical
manifolds by means of the IGE. Our final remarks appear in Section 6.

\section{The Information Geometry of Statistical Models}

The statistical models studied in \cite{cafarosoft} were a $3D$ uncorrelated
Gaussian statistical model and a $2D$ uncorrelated Gaussian statistical
model obtained from the higher-dimensional model via the introduction of an
additional information constraint that resembles the canonical minimum
uncertainty relation in quantum theory. For a brief and recent overview on
the IGAC, we refer to \cite{cafaroPD}. Note that the dimensionality ($2D$
and $3D$) pertains to the \textit{\ macroscopic} variables. Specifically,
the dimensionality of a curved statistical manifold equals the cardinality
of the set of time-varying statistical macro-variables necessary to
parametrize points on the manifold itself. Below, we examine the geometry of
the two-dimensional uncorrelated case, $2Du$, and then examine a new model,\
namely, the $2Dc$ model, where the microscopic variables are further
constrained by a \emph{covariance} term.

\subsection{The 2D Uncorrelated Model}

This section follows our previous work \cite{cafarosoft}, where we studied
the information geometry and the entropic dynamics of a $3D$ Gaussian
statistical model. We compared our analysis to that of a $2D$ Gaussian
statistical model obtained from the higher-dimensional model via the
introduction of an additional information constraint that resembled the
quantum mechanical canonical minimum uncertainty relation. We showed that
the chaoticity of the $2D$ Gaussian statistical model, quantified by means
of the Information Geometric Entropy and the Jacobi vector field intensity,
is softened with respect to the chaoticity of the $3D$ Gaussian statistical
model. In view of the similarity between the information constraint on the
variances and the phase-space coarse-graining imposed by the Heisenberg
uncertainty relations, we suggested that this provides a possible way of
explaining the phenomenon of the suppression of classical chaos operated by
quantization. The constraints on the \textit{microscopic} variables, $x$ and 
$y$, are: 
\begin{eqnarray}
\int p(x,y)ydxdy &=&0  \notag \\
\int p(x,y)xdxdy &=&\mu _{x}  \notag \\
&& \\
\int p(x,y)\left( x-\mu _{x}\right) ^{2}dxdy &=&\sigma _{x}^{2}  \notag \\
\int p(x,y)y^{2}dxdy &=&\sigma _{y}^{2}  \notag  \label{OC}
\end{eqnarray}

When these constraints are applied to the system, we use the method of
maximum (relative) \linebreak entropy \cite{AThesis} to obtain a family of
probability distributions that characterize the $3D$ uncorrelated Gaussian
statistical model:%
\begin{equation}
p\left( x\text{, }y|\mu _{x}\text{, }\sigma _{x}\text{, }\sigma _{y}\right) =%
\frac{1}{2\pi \sigma _{x}\sigma _{y}}\exp \left[ -\frac{1}{2\sigma _{x}^{2}}%
\left( x-\mu _{x}\right) ^{2}-\frac{1}{2\sigma _{y}^{2}}y^{2}\right] \text{}
\label{3D}
\end{equation}%
with $\sigma _{x}$ and $\sigma _{y}$ in $%
\mathbb{R}
_{0}^{+}$ and $\mu _{x}$ in $%
\mathbb{R}
$. The Gaussian here is two-dimensional in its microscopic space $(x$, $y)$,
but three-dimensional in its macroscopic (contextual or conditionally, given
parameters) space $\left( \mu _{x}\text{, }\sigma _{x}\text{, }\sigma
_{y}\right)$. For the $2D$ uncorrelated case, the probability distributions, 
$p\left( x\text{, }y|\mu _{x}\text{, }\sigma \right)$, that characterize the
model are given by:%
\begin{equation}
p\left( x\text{, }y|\mu _{x}\text{, }\sigma \right) \overset{\text{def}}{=}%
\frac{1}{2\pi \Sigma ^{2}}\exp \left[ -\frac{1}{2\sigma ^{2}}\left( x-\mu
_{x}\right) ^{2}-\frac{\sigma ^{2}}{2\Sigma ^{4}}y^{2}\right] \text{}
\label{GP}
\end{equation}%
with $\sigma$ in $%
\mathbb{R}
_{0}^{+}$ and $\mu _{x}$ in $%
\mathbb{R}
$. The probability distributions Equation (\ref{GP}) may be obtained from
Equation (\ref{3D}) with the addition of the following \emph{macroscopic
constraint}:%
\begin{equation}
\sigma _{x}\sigma _{y}=\Sigma ^{2}\text{}  \label{MC}
\end{equation}%
where $\Sigma ^{2}$ is a constant belonging to $%
\mathbb{R}
_{0}^{+}$ and $\sigma _{x}\equiv \sigma$. The macroscopic constraint
Equation (\ref{MC}) was chosen originally, because it resembles the quantum
mechanical canonical minimum uncertainty relation \linebreak where $x$
denotes the position of a particle and $y$, its conjugate momentum. Indeed,
in view of the \linebreak similarity between the constraint on the variances
Equation (\ref{MC}) and the phase-space coarse-graining imposed by the
Heisenberg uncertainty relations \cite{peres}, we seek a possible way of
explaining the phenomenon of the suppression of classical chaos when
operated by quantization within an information geometric framework.

We then relaxed the conditionality on the microscopic space to explore the
space of Gaussians described by $\mu _{x}$ and $\sigma$. The infinitesimal
Fisher-Rao line element, $ds_{2Du}^{2}$, for this model reads:%
\begin{equation}
ds_{2Du}^{2}=g_{lm}^{\left( 2Du\right) }\left( \theta \right) d\theta
^{l}d\theta ^{m}=\frac{1}{\sigma ^{2}}d\mu _{x}^{2}+\frac{4}{\sigma ^{2}}%
d\sigma ^{2}\text{}  \label{lineu}
\end{equation}%
where the Fisher-Rao information metric, $g_{lm}^{\left( 2Du\right) }\left(
\theta \right)$, is defined as \cite{amari}:%
\begin{equation}
g_{lm}^{\left( 2Du\right) }\left( \theta \right) \overset{\text{def}}{=}\int
dxdyp\left( x\text{, }y|\mu _{x}\text{, }\sigma \right) \frac{\partial \log
p\left( x\text{, }y|\mu _{x}\text{, }\sigma \right) }{\partial \theta ^{l}}%
\frac{\partial \log p\left( x\text{, }y|\mu _{x}\text{, }\sigma \right) }{%
\partial \theta ^{m}}\text{}
\end{equation}%
with $\theta \equiv \left( \theta ^{1}\text{, }\theta ^{2}\right) \overset{%
\text{def}}{=}\left( \mu _{x}\text{, }\sigma \right)$. Using Equation (\ref%
{lineu}), it follows that the non-vanishing connection coefficients, $\Gamma
_{ij}^{k}$, are given by:%
\begin{equation}
\Gamma _{12}^{1}=\Gamma _{21}^{1}=-\frac{1}{\sigma }\text{, }\Gamma
_{11}^{2}=\frac{1}{4\sigma }\text{, }\Gamma _{22}^{2}=-\frac{1}{\sigma }%
\text{}  \label{cristu}
\end{equation}%
The scalar curvature, $\mathcal{R}^{\left( 2Du\right)}$, of the probability
distributions in Equation (\ref{GP}) is given by:%
\begin{equation}
\mathcal{R}^{\left( 2Du\right) }=g^{11}\left( \theta \right)
R_{11}+g^{22}\left( \theta \right) R_{22}=-\frac{1}{2}\text{}
\label{riccis2}
\end{equation}%
with $g^{lm}g_{mk}=\delta _{k}^{l}$ and where the only non-vanishing Ricci
curvature tensor components, $R_{ij}$, are:%
\begin{equation}
R_{11}=-\frac{1}{4\sigma ^{2}}\text{, }R_{22}=-\frac{1}{\sigma ^{2}}\text{}
\end{equation}%
The sectional curvature \cite{Peterson}\ is independent of the tangent plane
chosen on any point of the manifold\ and is therefore constant, with value:%
\begin{equation}
\mathcal{K}^{\left( 2Du\right) }=-\frac{1}{4}\text{}
\end{equation}

As shown in \cite{cafarosoft}, the scalar curvature for the uncorrelated $3D$
case was $\mathcal{R}^{\left( 3Du\right) }=-1$. This implies that the $3D$
uncorrelated statistical model is \emph{globally} more negatively curved
than the $2D$ uncorrelated statistical model. This suggests that the $3D$
model might exhibit more chaotic features than the \linebreak $2D$ model.

\subsection{The 2D Correlated Model}

In addition to the original $3D$ uncorrelated Gaussian statistical model
constraints in Equation (\ref{OC}), we now add the following \emph{covariance%
} constraint to the model: 
\begin{equation}
\int p\left( x\text{, }y\right) xydxdy=\sigma _{xy}  \label{CovC}
\end{equation}

The probability distributions\ that characterize the $3D$ \emph{correlated}
Gaussian statistical model \linebreak Equation (\ref{3D}) now read:%
\begin{eqnarray}
&&p\left( x\text{, }y|\mu _{x}\text{, }\sigma _{x}\text{, }\sigma _{y}\text{,%
}~\sigma _{xy}\right) =  \notag  \label{Gaussian_cov} \\
&&\dfrac{1}{2\pi \sqrt{\sigma _{x}^{2}\sigma _{y}^{2}-\sigma _{xy}^{2}}}\exp
\left\{ \frac{-1}{2\left( \sigma _{x}^{2}\sigma _{y}^{2}-\sigma
_{xy}^{2}\right) }\left[ \left( x-\mu _{x}\right) ^{2}\sigma
_{y}^{2}+y^{2}\sigma _{x}^{2}-2\left( x-\mu _{x}\right) y\sigma _{xy}\right]
\right\}
\end{eqnarray}

Making a change of variable with regard to the covariance, $\sigma
_{xy}=r\sigma _{x}\sigma _{y}$, we obtain the standard bivariate normal
distribution, where the parameter, $r$, is the correlation coefficient
between $x$ and $y$ and assumes values within the ranges $-1\leq r\leq 1$.
Applying the macroscopic constraint Equation (\ref{MC}) to the covariance
constraint yields $\sigma _{xy}=r\Sigma ^{2}$, where $\Sigma ^{2}=\sigma
_{x}\sigma _{y}$ is a constant belonging to $%
\mathbb{R}
_{0}^{+}$ and $\sigma _{x}\equiv \sigma$. The $3D$ correlated model Equation
(\ref{Gaussian_cov}) now becomes: 
\begin{equation}
p\left( x\text{, }y|\mu _{x}\text{, }\sigma ,~r\right) \overset{\text{def}}{=%
}\dfrac{1}{2\pi \Sigma ^{2}\sqrt{1-r^{2}}}\exp \left\{ \frac{-1}{2\left(
1-r^{2}\right) }\left[ \frac{\left( x-\mu _{x}\right) ^{2}}{\sigma ^{2}}+%
\frac{y^{2}\sigma ^{2}}{\Sigma ^{4}}-\frac{2r\left( x-\mu _{x}\right) y}{%
\Sigma ^{2}}\right] \right\}  \label{CP}
\end{equation}

Assuming that the correlation between $x$ and $y$ is constant, we then relax
the conditionality on the microscopic space to explore the space of
Gaussians described by $\mu _{x}$ and $\sigma$ only. Thus, the infinitesimal
Fisher-Rao line element, $ds_{2Dc}^{2}$, reads: 
\begin{equation}
ds_{2Dc}^{2}=g_{lm}^{\left( 2Dc\right) }\left( \theta \right) d\theta
^{l}d\theta ^{m}=\frac{1}{\sigma ^{2}\left( 1-r^{2}\right) }d\mu _{x}^{2}+%
\frac{4}{\sigma ^{2}\left( 1-r^{2}\right) }d\sigma ^{2}\text{}  \label{linec}
\end{equation}%
where $\theta \equiv \left( \theta ^{1}\text{, }\theta ^{2}\right) \overset{%
\text{def}}{=}\left( \mu _{x}\text{, }\sigma \right)$. Observe that line
element Equation (\ref{linec}) is only valid provided \linebreak $\left(
1-r^{2}\right) > 0.$ Using Equation (\ref{linec}), it follows that the
non-vanishing connection coefficients, $\Gamma _{ij}^{k}$, are given by:%
\begin{equation}
\Gamma _{12}^{1}=-\frac{1}{\sigma }\text{, }\Gamma _{11}^{2}=\frac{1}{%
4\sigma },~\Gamma _{21}^{1}=-\frac{1}{\sigma },\text{~}\Gamma _{22}^{2}=-%
\frac{1}{\sigma }  \label{christc}
\end{equation}

The Ricci scalar curvature, $\mathcal{R}^{\left( 2Dc\right)}$, of the
probability distributions in Equation (\ref{CP}) is given by:%
\begin{equation}
\mathcal{R}^{\left( 2Dc\right) }=g^{11}\left( \theta \right)
R_{11}+g^{22}\left( \theta \right) R_{22}=-\frac{1}{2}+\frac{r^{2}}{2}\text{}
\label{riccis2c}
\end{equation}%
with $g^{lm}g_{mk}=\delta _{k}^{l}$. The only non-vanishing Ricci curvature
tensor components, $R_{ij}$, are:%
\begin{equation}
R_{11}=-\frac{1}{4\sigma ^{2}}\text{ and, }R_{22}=-\frac{1}{\sigma ^{2}}%
\text{}
\end{equation}

As in the previous case, the sectional curvature is independent of the
tangent plane chosen on any point of the manifold and is therefore constant,
with value:%
\begin{equation}
\mathcal{K}^{\left( 2Dc\right) }=-\frac{1}{4}\left( 1-r^{2}\right)
\end{equation}

Notice that the Ricci scalar of the correlated $2D$ model, $\mathcal{R}%
^{\left( 2Dc\right)}$ in Equation (\ref{riccis2c}), is $r$-dependent,
\linebreak with $\mathcal{R}^{\left( 2Dc\right)}$ approaching zero as $r^{2}$
tends to unity, while in the limit, $r\rightarrow 0$, we recover the scalar
curvature of the $2D$ uncorrelated Gaussian model Equation (\ref{riccis2}).
By including the macroscopic constraint Equation (\ref{MC}) with the
correlation constraint\ in Equation (\ref{CovC}), we limit what $\Sigma ^{2}$
can be; $\sigma _{xy}/r=\Sigma ^{2}$ and since $r^{2}<1$,%
\begin{equation}
\sigma _{xy}^{2}/\Sigma ^{4}<1~\text{, and, therefore, }\sigma _{xy}<\Sigma
^{2}\text{}
\end{equation}

Moreover, observe that the sectional curvature is correlation-dependent,
while the covariant Ricci tensor components are identical in both $2D$ cases.

\section{Geodesic Motion on Curved Statistical Manifolds}

In this section, we present the geodesic paths on the curved statistical
manifolds underlying the entropic dynamics of both the two-dimensional
correlated and uncorrelated Gaussian statistical models. Such paths are
obtained by integrating the geodesic equations given by \cite{landau}:%
\begin{equation}
\frac{d^{2}\theta ^{k}}{d\tau ^{2}}+\Gamma _{lm}^{k}\left( \theta \right) 
\frac{d\theta ^{l}}{d\tau }\frac{d\theta ^{m}}{d\tau }=0\text{}  \label{GEE}
\end{equation}%
where $\Gamma _{lm}^{k}\left( \theta \right)$ are the connection
coefficients.

Substituting Equation (\ref{christc}) into Equation (\ref{GEE}), the set of
nonlinear and coupled ordinary differential equations in Equation (\ref{GEE}%
) reads:%
\begin{eqnarray}
0 &=&\frac{d^{2}\mu _{x}}{d\tau ^{2}}-\frac{2}{\sigma }\frac{d\mu _{x}}{%
d\tau }\frac{d\sigma }{d\tau }\text{}  \notag \\
&& \\
0 &=&\frac{d^{2}\sigma }{d\tau ^{2}}+\frac{\sigma ^{2}+1}{4\sigma }\left( 
\frac{d\mu _{x}}{d\tau }\right) ^{2}-\frac{1}{\sigma }\left( \frac{d\sigma }{%
d\tau }\right) ^{2}\text{}  \notag
\end{eqnarray}

A suitable family of geodesic paths fulfilling the geodesic equations above
is given by:%
\begin{equation}
\mu _{x}\left( \tau \right) =\frac{\left( \mu _{0}+2\sigma _{0}\right) \left[
1+\exp \left( 2\sigma _{0}\lambda _{+}\tau \right) \right] -4\sigma _{0}}{%
\text{ }1+\exp \left( 2\sigma _{0}\lambda _{+}\tau \right) }\text{}
\label{cac3}
\end{equation}%
and:%
\begin{equation}
\sigma \left( \tau \right) =\frac{2\sigma _{0}\exp \left( \sigma _{0}\lambda
_{+}\tau \right) }{1+\exp \left( 2\sigma _{0}\lambda _{+}\tau \right) }\text{%
}  \label{cac4}
\end{equation}%
where $\mu _{0}\overset{\text{def}}{=}\mu _{x}\left( 0\right)$, $\sigma _{0}%
\overset{\text{def}}{=}\sigma \left( 0\right)$ and $\lambda _{+}$ belongs to 
$%
\mathbb{R}
^{+}$ \cite{cafarosoft}.

\section{Information Geometric Entropy}

In this section, the chaotic properties of the information-constrained
(entropic) dynamics on the underlying curved statistical manifolds are
quantified by means of the IGE. We point out that a suitable indicator of
temporal complexity (chaoticity) within the IGAC framework is provided by
the IGE, which, in the general case, reads \cite{AMC}:%
\begin{equation}
\mathcal{S}_{\mathcal{M}_{s}}\left( \tau \right) \overset{\text{def}}{=}\log %
\left[ \lim_{\tau \rightarrow \infty }\frac{1}{\tau }\int_{0}^{\tau }\int_{%
\mathcal{D}_{\theta }^{\left( \text{geodesic}\right) }\left( \tau ^{\prime
}\right) }\rho _{\left( \mathcal{M}_{s}\text{, }g\right) }\left( \theta ^{1}%
\text{,..., }\theta ^{n}\right) d^{n}\theta d\tau ^{\prime }\right] \text{}
\label{IGE}
\end{equation}%
where $\rho _{\left( \mathcal{M}_{s}\text{, }g\right) }\left( \theta ^{1}%
\text{,..., }\theta ^{n}\right)$ is the so-called Fisher density and equals
the square root of the determinant of the metric tensor, $g_{lm}\left(
\theta \right)$:%
\begin{equation}
\rho _{\left( \mathcal{M}_{s}\text{, }g\right) }\left( \theta ^{1}\text{%
,..., }\theta ^{n}\right) \overset{\text{def}}{=}\sqrt{g\left( \left( \theta
^{1}\text{,..., }\theta ^{n}\right) \right) }\text{}
\end{equation}%
The subscript, $\mathcal{M}_{s}$, in Equation (\ref{IGE}) denotes the curved
statistical manifold underlying the entropic dynamics. The integration
space, $\mathcal{D}_{\theta }^{\left( \text{geodesic}\right) }\left( \tau
^{\prime }\right)$, in Equation (\ref{IGE}) is defined as follows:%
\begin{equation}
\mathcal{D}_{\theta }^{\left( \text{geodesic}\right) }\left( \tau ^{\prime
}\right) \overset{\text{def}}{=}\left\{ \theta \equiv \left( \theta ^{1}%
\text{,..., }\theta ^{n}\right) :\theta ^{k}\left( 0\right) \leq \theta
^{k}\leq \theta ^{k}\left( \tau ^{\prime }\right) \right\} \text{}
\label{is}
\end{equation}%
where $k=1$,.., $n$ and $\theta ^{k}\equiv \theta ^{k}\left( s\right)$, with 
$0\leq s\leq \tau ^{\prime}$, such that $\theta ^{k}\left( s\right)$
satisfies Equation (\ref{GEE}). The integration space, $\mathcal{D}_{\theta
}^{\left( \text{geodesic}\right) }\left( \tau ^{\prime }\right)$, in
Equation (\ref{is}) is an $n$-dimensional subspace of the whole (permitted)
parameter space, $\mathcal{D}_{\theta }^{\left( \text{tot}\right)}$. The
elements of $\mathcal{D}_{\theta }^{\left( \text{geodesic}\right) }\left(
\tau ^{\prime }\right)$ are the $n$-dimensional macro-variables, $\left\{
\theta \right\}$, whose components, $\theta ^{k}$, are bounded by the
specified limits of integration $\theta ^{k}\left( 0\right)$ and $\theta
^{k}\left( \tau ^{\prime }\right)$ with \mbox{$k=1$,..., $n$}. The limits of
integration are obtained via integration of the $n$-dimensional set of
coupled nonlinear second order ordinary differential equations
characterizing the geodesic equations. Formally, the IGE is defined in terms
of an averaged parametric $\left( n+1\right)$-fold integral ($\tau$ is the
parameter) over the multi-dimensional geodesic paths connecting $\theta
\left( 0\right)$ to $\theta \left( \tau \right)$.

In the cases being investigated, using Equations (\ref{cac3})--(\ref{IGE}),
it follows that the asymptotic expressions of the IGE for the uncorrelated
model, $\mathcal{S}_{\mathcal{M}_{s}}^{\left( 2Du\right)}$, and for the
correlated model, $\mathcal{S}_{\mathcal{M}_{s}}^{\left( 2Dc\right)}$,
become:%
\begin{equation}
\mathcal{S}_{\mathcal{M}_{s}}^{\left( 2Du\right) }\left( \tau \right) =\log 
\mathcal{V}_{\mathcal{M}_{s}}^{\left( 2Du\right) }\left( \tau \right) 
\overset{\tau \gg 1}{\approx }\sigma _{0}\lambda _{+}\tau \text{~and~}%
\mathcal{S}_{\mathcal{M}_{s}}^{\left( 2Dc\right) }=\log \mathcal{V}_{%
\mathcal{M}_{s}}^{\left( 2Dc\right) }\left( \tau \right) \overset{\tau \gg 1}%
{\approx }\sigma _{0}\lambda _{+}\tau  \label{IGE12}
\end{equation}%
since:%
\begin{eqnarray}
&&\mathcal{V}_{\mathcal{M}_{s}}^{\left( 2Du\right) }\left( \tau \right) 
\overset{\tau \gg 1}{\approx }\left[ \left( \frac{\mu _{0}+2\sigma _{0}}{%
\sigma _{0}^{2}\lambda _{+}}\right) \frac{\exp \left( \sigma _{0}\lambda
_{+}\tau \right) }{\tau }\right]  \notag  \label{IGCfinal} \\
&& \\
&&\mathcal{V}_{\mathcal{M}_{s}}^{\left( 2Dc\right) }\left( \tau \right) 
\overset{\tau \gg 1}{\approx }\frac{1}{\left( 1-r^{2}\right) }\mathcal{V}_{%
\mathcal{M}_{s}}^{\left( 2Du\right) }\left( \tau \right) \text{}  \notag
\end{eqnarray}%
From Equation (\ref{IGE12}), we observe that:%
\begin{equation}
\mathcal{S}_{\mathcal{M}_{s}}^{\left( 2Dc\right) }\overset{\tau \gg 1}{%
\approx }\mathcal{S}_{\mathcal{M}_{s}}^{\left( 2Du\right) }\text{}
\end{equation}

The IGE does not change asymptotically for either of the $2D$ models being
considered. Equation (\ref{IGCfinal}) \linebreak is quite interesting, since
it quantitatively shows that the information geometric complexity (IGC), $%
\mathcal{V}_{\mathcal{M}_{s}}^{\left( 2Dc\right)}$, of the correlated $2Dc$
model diverges as the correlation coefficient [introduced via the constraint
Equation (\ref{CovC})] approaches unity. As expected, the two cases are
identical for $r = 0$. In \cite{cafarosoft}, the IGE of the quantum-like
model $2Du$ was less than the IGE of the classical model ($3D$). That result
indicated a weaker (softer) chaoticity for the $2Du$ model. Here, the
comparison of the uncorrelated and correlated models shows that further
constraining the quantum-like model $2Du$ with a covariance term, $\sigma
_{xy}$, does not lead to any additional global softening.

\section{Jacobi Vector Field Intensity}

There seems to be no change to the chaoticity when we further constrain the
old, quantum-like model ($2Du$) with a covariance term, $\sigma _{xy}$,
since the IGE for each $2D$ model is identical in the asymptotic limit.
However, when considering chaoticity characterizations (geodesic spread), it
is the \emph{local} curvature of the manifold that must be examined. This
information is encoded in the sectional curvatures of the manifold when it
is isotropic (maximally symmetric). When the manifold is anisotropic
(non-maximally symmetric), the Riemann curvature tensor components come into
play.

From above, we see that, indeed, the sectional curvatures, $\mathcal{K}$, of
both $2D$ models are constant (and, consequently, maximally symmetric) and\
exhibit the relationship:%
\begin{equation}
\mathcal{K}^{\left( 2Dc\right) }=-\frac{1}{4}\left( 1-r^{2}\right) \geq -%
\frac{1}{4}=\mathcal{K}^{\left( 2Du\right) }\text{}  \label{sectional}
\end{equation}%
with $r^{2} < 1$. However, since the sectional curvatures are different, to
the extent that the $2Dc$ model depends on the correlation coefficient, the 
\emph{local} curvature is different. We can then follow \cite{cafarosoft} in
integrating the Jacobi-Levi-Civita (JLC) equation describing the geodesic
spread. Omitting technical details, we find that the asymptotic temporal
behavior of the Jacobi vector field intensities, $\mathcal{J}$, on such
maximally symmetric statistical manifolds satisfy the following inequality
relation, which is closely related to Equation (\ref{sectional}):%
\begin{equation}
\mathcal{J}^{\left( 2Du\right) }\left( \tau \right) \overset{\tau
\rightarrow \infty }{\approx }\exp \left( +\sqrt{\mathcal{K}^{\left(
2Du\right) }}\tau \right) \geq \exp \left( +\sqrt{\mathcal{K}^{\left(
2Dc\right) }}\tau \right) \overset{\tau \rightarrow \infty }{\approx }%
\mathcal{J}^{\left( 2Dc\right) }\left( \tau \right) \text{}
\end{equation}

This would imply that there is indeed a \emph{local softening} of the
geodesic spreads on the quantum-like model, the $2Du$ model, when it is
further constrained by a covariance constraint, $\sigma _{xy}$, manifested
in the correlation term, $r$.

\section{Other Considerations}

\begin{itemize}
\item While $\sigma _{xy}=r\sigma _{x}\sigma _{y}$\ is physically sensible
for $x$\ and $y$\, representing either position-position, \linebreak
momentum-momentum or position-momentum pairs, the macro-quantum-like
constraint \linebreak $\Sigma ^{2}=\sigma _{x}\sigma _{y}$\ with $\Sigma
^{2} $\ being constant and $x$\ and $y$\ representing either
position-position or momentum-momentum pairs is physically ambiguous in view
of the fact that all physical observables commute with the permutation
operator. This seems to suggest that the constraint $\Sigma ^{2}=\sigma
_{x}\sigma _{y}$\ is only active when $x$\ and $y$\ refer to micro-variables
that are not self-similar. Might one be able to use empirical data to trace
the information-theoretic conditions that either relax the constraint $%
\Sigma ^{2}=\sigma _{x}\sigma _{y}$\ (so that $\Sigma ^{2}\neq $constant) or
renders it active? It should be further noted that even if one considers
micro-variables with dissimilar dimensions, then while the covariance
constraint and the macro-quantum-like constraint would be compatible, one
would then have to consider a more general form of the macro-quantum-like
constraint Equation (\ref{MC})\, due to the presence of entanglement. This
will be considered for a future work.

\item We also stress that our information geometric analysis could
accommodate non-minimum uncertainty-like relations. However, such an
extension would require a more delicate analysis where maximum relative
entropy methods are used to process information in the presence of
inequality constraints \cite{ishwar}. However, for this, as well as the more
general uncertainty constraint, a deeper analysis is needed, and we leave
that for future investigations. Our work is especially relevant for the
quantification of soft chaos effects in entropic dynamical models used to
describe actual physical systems when only incomplete knowledge about them
is available \cite{AM}.

\item Statistical complexity is a quantity that measures the amount of
memory needed, on average, to statistically reproduce a given configuration 
\cite{jim}. In the same vein of our works in \cite{cafarosoft}, a recent
investigation claims that quantum mechanics can reduce the statistical
complexity of classical models \cite{nc}. Specifically, it was shown that
mathematical models featuring quantum effects can be as predictive as
classical models, although implemented by simulators that require less
memory, that is, less statistical complexity. Of course, these two works use
different definitions of complexity, and their ultimate goal is definitively
not the same. However, it is remarkable that both of them exploit some
quantum feature, Heisenberg's uncertainty principle in \cite{cafarosoft} and
the quantum state discrimination (information storage) method in \cite{nc},
to exhibit the complexity softening effects. Is there any link between
Heisenberg's uncertainty principle and quantum state discrimination?
Recently, it was shown that any violation of uncertainty relations in
quantum mechanics also leads to a violation of the second law of
thermodynamics \cite{ester}. In addition, it was reported in \cite{ralph}
that a violation of Heisenberg's uncertainty principle allows perfect
\linebreak state discrimination of non-orthogonal states, which, in turn,
violates the second law of thermodynamics \cite{peres}. The possibility of
distinguishing non-orthogonal states is directly related to the question of
how much information we can store in a quantum state. Information storage
and memory are key quantities for the characterization of statistical
complexity. In view of these considerations, it would be worthwhile to
explore the possible thermodynamic link underlying these two different
complexity measures \cite{carluccio}.
\end{itemize}

All these considerations will be the subject of forthcoming efforts.

\section{Conclusions}

In a previous paper \cite{cafarosoft}, we studied the information geometry
of an uncorrelated $3D$ Gaussian \mbox{statistical} model with an additional
information constraint resembling the canonical minimum \mbox{uncertainty}
relation, which, here, we called $2Du$. We showed that the chaoticity of
such a modified Gaussian statistical model (quantum-like model, $2Du$),
quantified by means of the Information Geometric Entropy \cite{cafaroPD} and
the Jacobi vector field intensity, was indeed softened with respect to the
chaoticity of the standard Gaussian statistical model (classical-like model, 
$3D$). However, the statistical model was limited in that we assumed there
was no correlation between the constituents of the phase~space.

In this paper, we expanded our previous findings by further constraining the
quantum-like $2Du$ model with a covariance term, $\sigma _{xy}$. It was
shown that the Ricci scalars of the two $2D$ models, Equations (\ref{riccis2}%
) and (\ref{riccis2c})\, varied by a constant related to this covariance
term, Equation (\ref{CovC}). It seems that the IGE is insensitive to the
presence of correlation terms, since the asymptotic behavior of the IGE of
the two $2D$\ models, $S_{\mathcal{M}_{s}}^{\left( 2Du\right) }\left( \tau
\right)$\ and $S_{\mathcal{M}_{s}}^{\left( 2Dc\right) }\left( \tau \right)$%
,\ are identical. Although the IGE\ analysis seemed to indicate that there
was no further global softening, examination of the Jacobi vector field
intensity seems to indicate that the softening only appears locally
(geodesic spread-deviation equations). Therefore, when the quantum-like $2Du$
model is further constrained by the knowledge of a covariance term, $\sigma
_{xy} $, no softening appears at the \emph{global }(geodesic equations)
scale, but only appears \emph{\ locally}, where this softness is dependent
on the correlation term, $r$. In a forthcoming investigation, we will
consider $\tau$-dependent $r$ quantities and study whether or not this
formal identical behavior between the IGEs is preserved for such cases, as
well. The stronger condition of $\tau$-dependent $r$ may affect the
chaoticity features of the correlation constrained $2Dc$ model at a global
scale.

Finally, we would like to point out a very intriguing analogy inspired by
one of the referees (which we suspect to be very profound) between
Einstein's Equivalence Principle \cite{felice} and the local softening
effect considered in this work. Einstein's Equivalence Principle states that
gravitation, like space-time curvature, works only globally, while locally,
there is no gravitational field: physics is simply connected only locally.
It may be worthwhile deepening this point also in future investigations,
taking into proper consideration the fact that while Einstein was discussing
space-time regions, our considerations concern regions on curved statistical
manifolds.

\begin{acknowledgments}
We acknowledge that an earlier version of this work was presented at MAXENT
2012 : International Workshop on Bayesian Inference and Maximum Entropy
Methods in Science and Engineering held at the Max-Planck-Institut fur
Plasmaphysik (IPP) in Garching bei Munchen, Germany. We would also like to
thank the referees for inspiring our final comment and giving us additional
insight into our work.
\end{acknowledgments}


\begin{thebibliography}{99}
\bibitem{caron1} Caron, L.A.; Jiraria, H.; Krögera, H.; Luob, X.Q.;
Melkonyana, G.; Moriartyd, K.J.M. \linebreak Quantum chaos at finite
temperature. \emph{Phys. Lett. A} \textbf{2001}, \emph{288}, 145--153.

\bibitem{peres} Peres, A. \emph{Quantum Theory: Concepts and Methods};
Volume 57, fundamental theories of physics; Springer: New York, NY, USA,
1995.

\bibitem{cafarosoft} C. Cafaro, A. Giffin, C. Lupo, and S. Mancini, \emph{%
Softening the complexity of entropic motion on curved statistical manifolds}%
, Open Systems \& Information Dynamics \textbf{19}, 1250001 (2012).

\bibitem{cafaroPD} Cafaro, C.; Mancini, S. Quantifying the complexity of
geodesic paths on curved statistical manifolds through information geometric
entropies and Jacobi fields. \emph{Physica D } \textbf{2011}, \emph{240},
607--618.

\bibitem{catichaED} Caticha, A. Entropic dynamics. \emph{AIP Conf. Proc.} 
\textbf{2002}, \emph{617}, 302--313.

\bibitem{amari} Amari, S.; Nagaoka, H. \emph{Methods of Information Geometry}%
; Oxford University Press: Oxford, UK, 2000.

\bibitem{AThesis} Giffin, A. Maximum entropy: The universal method for
inference. Ph.D. Thesis, State University of New York, Albany, NY, USA, 2008.

\bibitem{c2-1} Shore, J.E.; Johnson, R.W. Axiomatic derivation of the
principle of maximum entropy and the principle of minimum cross-entropy. 
\emph{IEEE Trans. Inf. Theory} \textbf{1980}, \emph{26}, 26--37.

\bibitem{c2-2} Shore, J.E.; Johnson, R.W. Properties of cross-entropy
minimization. \emph{IEEE Trans. Inf. Theory} \textbf{1981}, \emph{27},
472--482.

\bibitem{c4} Caticha, A.; Giffin, A. Updating probabilities. \emph{AIP Conf.
Proc.} \textbf{2006}, \emph{872}, 31--42.

\bibitem{cafarothesis} Cafaro, C. The Information Geometry of Chaos. Ph.D.
Thesis, State University of New York, Albany, NY, USA, 2008.

\bibitem{cafaroCSF} Cafaro, C. Works on an information geometrodynamical
approach to chaos. \emph{Chaos Solitons \& Fractals} \textbf{2009}, \emph{41}%
, 886--891.

\bibitem{Bubhardt} Busshardt, M.; Freyberger, M. Decoherent dynamics of two
nonclassically correlated particles. \textit{Phys. Rev. A} \textbf{2007}, 
\emph{75}, 052101.

\bibitem{KACM} Kim, D.-H.; Ali, S.A.; Cafaro, C.; Mancini, S. Information
geometry of quantum entangled Gaussian wave-packets. \emph{Physica A} 
\textbf{2012}, \emph{391}, 4517--4556.

\bibitem{Serafini} Serafini, A.; Adesso, G. Standard forms and entanglement
engineering of multimode Gaussian states under local operations. \textit{J.
Phys. A: Math. Theor. } \textbf{2007}, \textit{40},
doi:10.1088/1751-8113/\linebreak 40/28/S13.

\bibitem{peter} Braunstein, S.; van Loock, P. Quantum information with
continuous variables. \textit{Rev. Mod. Phys.} \textbf{2005}, \textit{77},
513--577.

\bibitem{Peterson} Petersen, P. \emph{Riemannian Geometry}; Springer:
Berlin, Germany, 2006.

\bibitem{landau} Landau, L.D.; Lifshitz, E.M. \emph{The Classical Theory of
Fields}; Pergamon: London, UK, 1962.

\bibitem{AMC} Cafaro, C.; Giffin, A.; Ali, S.A.; Kim, D.-H. Reexamination of
an information geometric construction of entropic indicators of complexity. 
\textit{Appl. Math. Comput.} \textbf{2010}, \textit{217}, 2944--2951.

\bibitem{ishwar} Ishwar, P.; Moulin, P. On the existence and
characterization of the maxent distribution under general moment inequality
constraints. \textit{IEEE Trans. Inf. Theory} \textbf{2005}, \textit{51},
3322--3333.

\bibitem{AM} Peng, L.; Sun, H.; Sun, D.; Yi, J. The geometric structures and
instability of entropic dynamical models. \textit{Adv. Math.} \textbf{2011}, 
\textit{227}, 459--471.

\bibitem{jim} Crutchfield, J.P.; Young, K. Inferring statistical complexity. 
\textit{Phys. Rev. Lett.} \textbf{1989}, \textit{63}, 105--108.

\bibitem{nc} Gu, M.; Cui, J.; Kwek, L.C.; Santos, M.F.; Fan, H.; Vedral, V.
Quantum phases with differing computational power. \textit{Nat. Commun.} 
\textbf{2012}, \textit{3}, doi:10.1038/ncomms1809.

\bibitem{ester} Hanggi, E.; Wehner, S. A violation of the uncertainty
principle implies a violation of the second law of thermodynamics. \textbf{%
2012}, arXiv:quant-ph/1205.6894.

\bibitem{ralph} Pienaar, J.L.; Ralph, T.C.; Myers, C.R. Open timelike curves
violate Heisenberg's uncertainty principle. \textbf{2012},
arXiv:quant-ph/1206.5485.

\bibitem{carluccio} Cafaro, C. Information Geometric Complexity of Entropic
Motion on Curved Statistical Manifolds, \textit{Proceedings of the 12th
Joint European Thermodynamics Conference, JETC 2013}, Eds. M. Pilotelli and
G.P. Beretta (ISBN 978-88-89252-22-2, Snoopy, Brescia, Italy, 2013), pp.
110--118.

\bibitem{felice} De Felice, F.; Clarke, J.S. \emph{Relativity on Curved
Manifolds}; Cambridge University Press: Cambridge, UK, 1990.
\end{thebibliography}
\end{document}